\documentclass[showpacs,prl,amsmath,amssymb,superscriptaddress,nobibnotes,twocolumn, preprintnumbers]{revtex4-1}
\usepackage{graphicx}
\usepackage{endnotes}
\usepackage{bm}
\usepackage{epsfig}
\usepackage{amsmath}

\newcommand{\ie}{{\it i.e.}}

\newcommand{\SrIr}{Sr$_{3}$Ir$_4$Sn$_{13}$}
\newcommand{\SrRh}{Sr$_{3}$Rh$_4$Sn$_{13}$}

\newcommand{\CaSrRhx}{(Ca$_{x}$Sr$_{1-x}$)$_3$Rh$_4$Sn$_{13}$}
\newcommand{\CaSrIrx}{(Ca$_{x}$Sr$_{1-x}$)$_3$Ir$_4$Sn$_{13}$}
\newcommand{\CaRh}{Ca$_{3}$Rh$_4$Sn$_{13}$}
\newcommand{\CaIr}{Ca$_{3}$Ir$_4$Sn$_{13}$}
\newcommand{\Tstar}{$T^*$}

\begin{document}

%%%%%%%%%%%%%%%%%%%%% TITLE %%%%%%%%%%%%%%%%%%%%

\title{Strong Coupling Superconductivity in the Vicinity of the Structural Quantum Critical Point in (Ca$_x$Sr$_{1-x}$)$_3$Rh$_4$Sn$_{13}$}
%%%%%%%%%%%%%%%%%%%% AUTHORS %%%%%%%%%%%%%%%%%%
\author{Wing Chi Yu}
\email{wcyu@phy.cuhk.edu.hk\\ skgoh@phy.cuhk.edu.hk}
\author{Yiu Wing Cheung}
\affiliation{Department of Physics, The Chinese University of Hong Kong, Shatin, New Territories, Hong Kong, China}

\author{Paul J. Saines}
\affiliation{Department of Chemistry, University of Oxford, Inorganic Chemistry Laboratory, South Parks Road, Oxford OX1 3QR, United Kingdom}

\author{Masaki Imai}
\author{Takuya Matsumoto}
\author{Chishiro Michioka}
\affiliation{Department of Chemistry, Graduate School of Science, Kyoto University, Kyoto 606-8502, Japan}
\author{Kazuyoshi Yoshimura}
\affiliation{Department of Chemistry, Graduate School of Science, Kyoto University, Kyoto 606-8502, Japan}
\affiliation{Research Center for Low Temperature and Materials Sciences, Kyoto University, Kyoto 606-8501, Japan}

\author{Swee K. Goh}
\email{wcyu@phy.cuhk.edu.hk\\ skgoh@phy.cuhk.edu.hk}
\affiliation{Department of Physics, The Chinese University of Hong Kong, Shatin, New Territories, Hong Kong, China}
\date{\today}
%\preprint{Resub ver. 4.0}
%%%%%%%%%%%%%%%%%%% ABSTRACT %%%%%%%%%%%%%%%%%%%%

\begin{abstract}
The family of the superconducting quasi-skutterudites \CaSrRhx\ features a structural quantum critical point at $x_c=0.9$, around which a dome-shaped variation of the superconducting transition temperature $T_c$ is found. Using specific heat, we probe the normal and the superconducting states of the entire series straddling the quantum critical point. Our analysis indicates a significant lowering of the effective Debye temperature on approaching $x_c$, which we interpret as a result of phonon softening accompanying the structural instability. Furthermore, a remarkably large enhancement of $2\Delta/k_BT_c$ and $\Delta C/\gamma T_c$ beyond the Bardeen-Cooper-Schrieffer (BCS) values is found in the vicinity of the structural quantum critical point. The phase diagram of (Ca$_{x}$Sr$_{1-x}$)$_3$Rh$_4$Sn$_{13}$ thus provides a model system to study the interplay between structural quantum criticality and strong electron-phonon coupling superconductivity.
\end{abstract}

%74.25.-q - properties of superconductors
%62.50.-p - high pressure effects in solids and liquids
%74.25.Op - superconductors - critical fields.
%74.40.Kb	Quantum critical phenomena
%74.62.-c	Transition temperature variations, phase diagrams
%65.40.Ba	Heat capacity
%64.70.Tg	Quantum phase transitions

\maketitle

%%%%%%%%%%%%%%%%%%%%% INTRO %%%%%%%%%%%%%%%%%%%%
%\section{Introduction}

The interplay between superconductivity and structural instability has been an important theme in condensed matter physics. The proximity to structural instability has been suggested to play a role in several superconductors with relatively high transition temperature ($T_c$) \cite{Testardi1957,Hinks1988,Shanks1974, Kim2006, Gauzzi2007,Weller2005,Emery2005}, with recent examples including IrTe$_2$ \cite{Yang2012,Pyon2012,Fang2013} as well as Fe- and Ni-based superconductors \cite{Cruz2008, Yoshi2012, Niedziela2011, Kudo2012, Hirai2012}. To examine the prospect of stabilising or even enhancing superconductivity near structural instability, access to superconducting systems featuring a second-order structural phase transition, which can be tuned away, is highly desirable.

A large family of superconducting stannides \cite{Remeika1980, Espinosa1980} with composition A$_3$T$_4$Sn$_{13}$, where A=La,Sr,Ca and T=Rh,Ir, has recently attracted considerable attention \cite{Yang10,Kase11, Hayamizu11, Wang2012,Klintberg2012,Zhou2012,Gerber2013,Liu13,Slebarski14,Tompsett14,
Kuo14,Biswas14,Sarkar14,Fang14,Goh14,XChen15}. $T_c$ in these compounds are typically less than 10~K, and the superconducting gap function is established to be nodeless at the Fermi surface \cite{Kase11,Hayamizu11,Wang2012,Zhou2012}. What makes these systems interesting is the discovery of an additional second-order structural phase transition at \Tstar\ in some of the compositions \cite{Kuo14, Klintberg2012, Goh14}. Systematic variation of \Tstar\ has been observed when the unit cell volume of the crystal is varied via chemical substitution or applied pressure, resulting in phase diagrams suggestive of a strong interplay between the structural and superconducting orders, which is in striking resemblance to the phase diagrams of heavily studied superconductors found on the border of magnetism
\cite{Mathur1998,Gegenwart2008,Paglione2010,Ishida2009,Hashimoto2012,Shibauchi2014}.

\SrIr\ and \SrRh\ are two prominent members of the family which exhibit both $T^*$ and $T_c$. Clear lambda-like jump in the specific heat can be seen at $T^*=147$~K and 138~K in \SrIr\ \cite{Kuo14} and \SrRh \cite{Goh14}, respectively. Crucially, \Tstar\ can be suppressed rapidly by applying pressure, or substituting calcium by strontium \cite{Klintberg2012,Goh14}. In the Ir substitution series, \ie\ \CaSrIrx, $T^*$ reaches as low as 33~K when $x=1$. Full suppression of \Tstar\ can only be achieved by applying 18~kbar to \CaIr \cite{Klintberg2012}. In the Rh substitution series, \Tstar\ can be fully suppressed solely with calcium substitution, giving rise to a structural quantum critical point (QCP) at ambient pressure at $x\approx0.9$ \cite{Goh14}. Concurrent with the suppression of $T^*$, $T_c$ rises gently in both Ir- and Rh-series, and reaches the maximum value near where $T^*\rightarrow0$ \cite{Klintberg2012, Goh14}.

A second-order structural phase transition involves the softening of the phonon mode that is responsible for the structural order \cite{Cowley1980, Dovebook}. This will inevitably affect the low energy excitations of the system. In addition, the nodeless nature of the superconducting gap function suggests a conventional Cooper pairing mechanism mediated by phonons, allowing the discussion of the coupling strength. The tunability of the \CaSrRhx\
series across the structural QCP without applying high pressure offers an important opportunity to understand the underlying lattice dynamics and the relationship between structural and superconducting orders. In this Letter, we report the results of our heat capacity measurements on \CaSrRhx\ for seven Ca concentrations, spanning the range from $x=0$ to $x=1$. The data enable us to extract important parameters relevant to both superconducting and structural phase transitions, which reveal a remarkable trend on approaching the structural QCP.

%%%%%%%%%%%%%%%%% EXPERIMENTAL %%%%%%%%%%%%%%%%%%%%
%\section{Experimental}

Single crystals of (Ca$_x$Sr$_{1-x}$)$_3$Rh$_4$Sn$_{13}$ were synthesized by Sn flux method using similar parameters as described in ref. \cite{Yang10}. Heat capacity was measured in a Physical Property Measurement System (Quantum Design) using a standard pulse relaxation method. The background heat capacity of the sample platform and Apiezon N-grease, which was used as adhesive, was carefully measured from room temperature down to the lowest attainable temperature. To accurately determine the normal state heat capacity of the samples, additional background measurements were performed at high magnetic field ($\leq$~5~T). The mass of the samples ranges from $\sim$5~mg to $\sim$50~mg.

%%%%%%%%%%%%%%%%%%%%% RESULTS %%%%%%%%%%%%%%%%%%
%\section{Results and Discussion}
%%%%%%%%%%%%%%%%Figure 1
\begin{figure}[!t]\centering
      \resizebox{8cm}{!}{
              \includegraphics{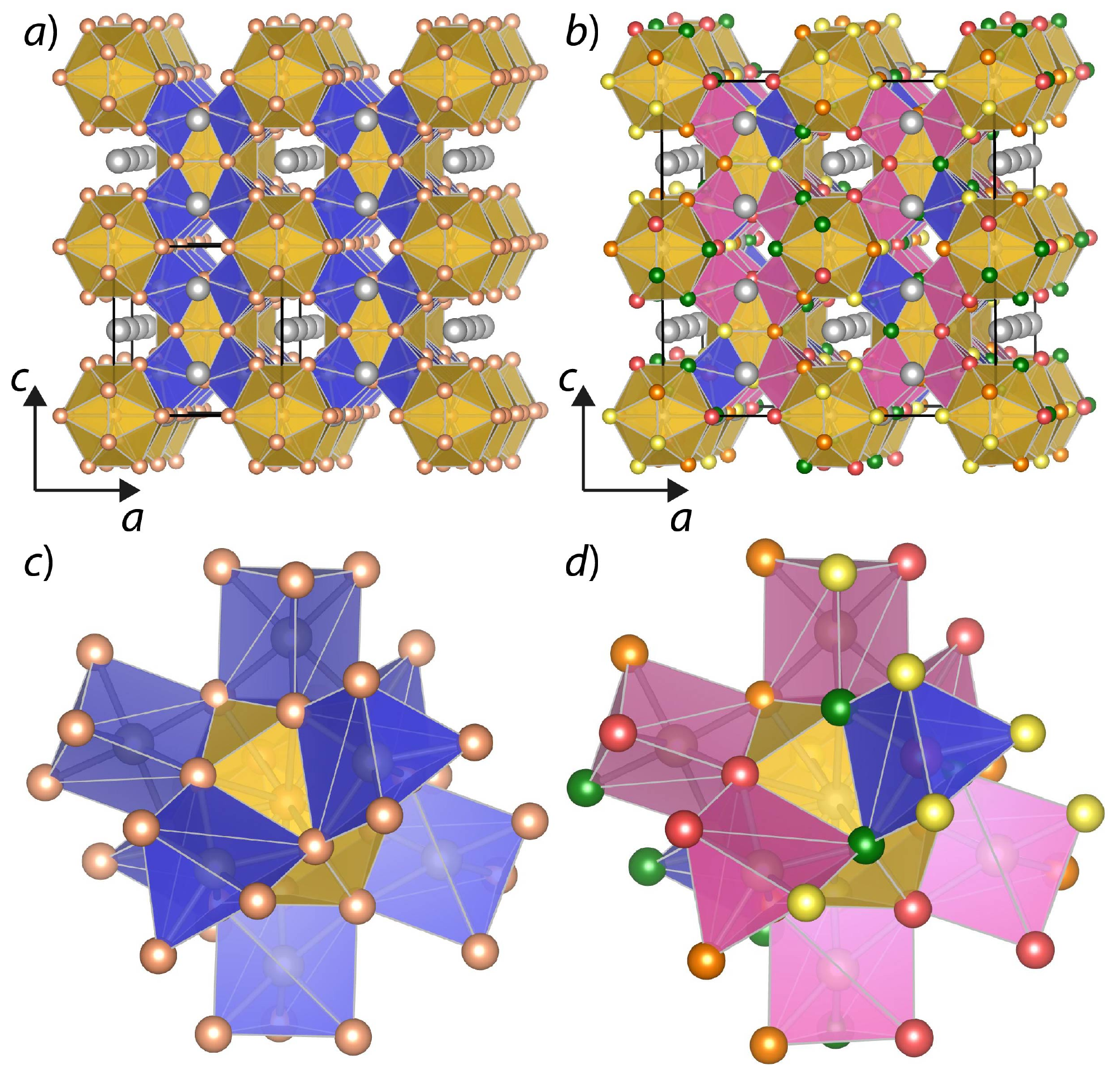}}                				
              \caption{\label{fig1} (Color online) The structure of \SrRh\ (a) above and (b) below \Tstar. The Sr, Rh, and Sn atoms are represented by spheres with descending sizes, and icosahedra, untilted and tilted trigonal prisms are coloured in tangerine, blue and pink. Crystallographically distinct Sn(2) atoms are coloured green, red, yellow and orange, and a single unit cell of each structure is indicated. (c) and (d) show the subtle tilting of the trigonal prisms RhSn(2)$_{6}$ around an Sn(1)Sn(2)$_{12}$ icosahedron.}
\end{figure}

Above \Tstar, \SrRh\ adopts a cubic $Pm\bar{3}n$ structure (Fig. 1(a)) featuring Sn(1)Sn(2)$_{12}$ icosahedra connected via RhSn(2)$_6$ trigonal prisms, with the Sr cations occupying the space between icosahedra. In this high temperature centrosymmetric phase, the bond lengths in the Sn(1)Sn(2)$_{12}$ icosahedra are identical. Below \Tstar, these icosahedra distort leading to each having four groups of Sn(1)-Sn(2) bonds of different lengths \cite{Goh14}. This requires six of the eight trigonal prisms around an icosahedra to tilt, with only those along the [111] direction remaining untilted. This results in a symmetry lowering to a non-centrosymmetric $I\bar{4}3d$ cubic structure with a doubled unit cell (see Fig. 1(b)).

%%%%%%%%%%%%%%%%Figure 2
\begin{figure}[!t]\centering
       \resizebox{8.5cm}{!}{
              \includegraphics{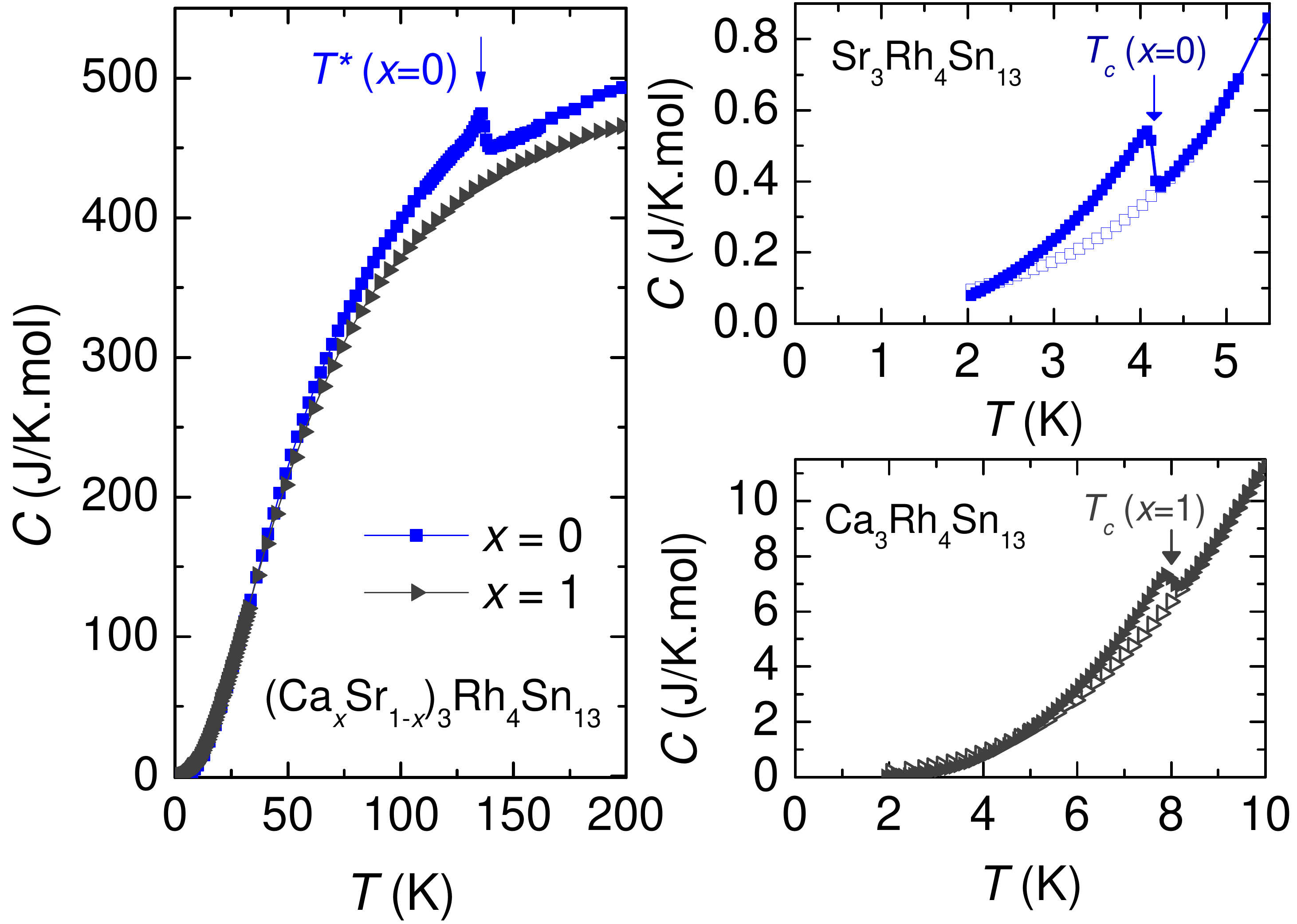}}                				
              \caption{\label{fig2} (Color online) (Left) The specific heat of \SrRh\ ($x=0$) and \CaRh\ ($x=1$) as a function of temperature at zero external magnetic field (closed symbols). \Tstar\ indicates the temperature at which a structural transition takes place. (Right) A closeup of the specific heat at low temperatures. The midpoint of the jump defines $T_c$, which can be suppressed by applying external magnetic field (open symbols).
              %The specific heat of \SrRh\ ($x=0$) and \CaRh\ ($x=1$) as a function of temperature at zero external magnetic field. \Tstar\ indicates the temperature at which a structural transition takes place. Inset shows a closeup of the specific heat at low temperature. The midpoint of the cusp in the specific heat defines the superconducting transition temperature $T_c$.
              }
\end{figure}
%%%%%%%%%%%%%%%%%%%%
Fig. \ref{fig2} shows the temperature dependence of the specific heat for \SrRh\ and \CaRh. In \SrRh, a lambda-like jump in the specific heat occurs at \Tstar$\sim$138~K, which corresponds to a second-order structural phase transition \cite{Goh14}. At low temperatures, another jump in the specific heat is detected (see the right panel of Fig. \ref{fig2}), which marks $T_c$ and is in excellent agreement with previous electrical resistivity measurement \cite{Goh14} (see Fig. S1 of Supplemental Material). In \CaRh, the specific heat varies smoothly except at $T_c=8$~K. This is again consistent with electrical resistivity measurement which did not detect any evidence of \Tstar\ (Fig. S1).

\begin{figure}[!t]\centering
       \resizebox{8cm}{!}{
              \includegraphics{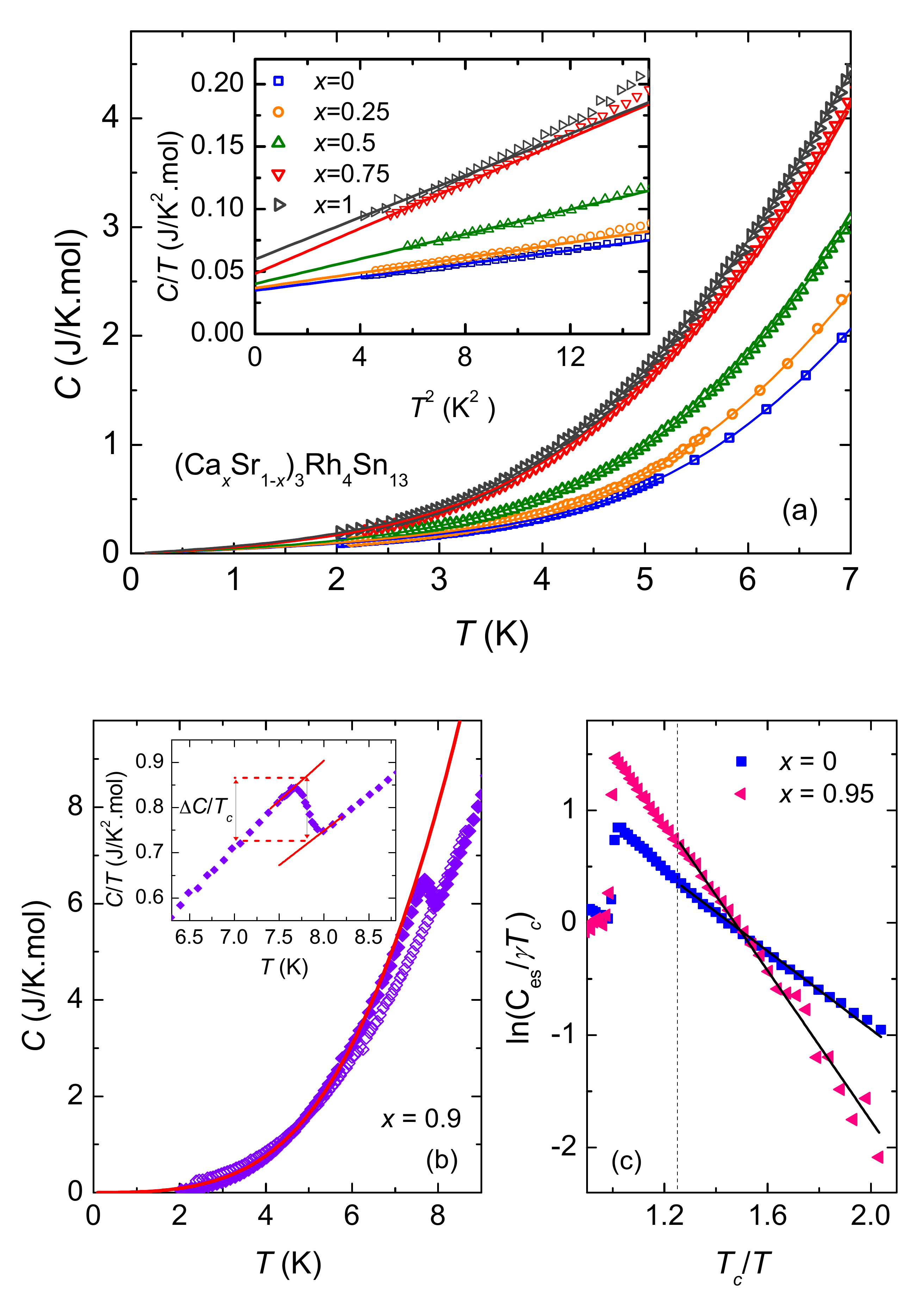}}              				
              \caption{\label{fig3} (Color online) (a) Temperature dependence of the specific heat of \CaSrRhx\ in the normal state for $x=0,0.25,0.5,0.75,1$. Inset shows $C/T$ as a function of $T^2$. The experimental data (open symbols) at low temperatures are fitted linearly (solid lines). $\gamma$ and $\beta$ in Eq. (\ref{eq:Cn}) are extracted from the intercept and slope of the fit, respectively. In the main panel, the measured specific heat is fitted using the full expression of Eq. (\ref{eq:Cn}) with the extracted values of $\gamma$ and $\beta$. (b) The specific heat of the critical composition ($x=0.9$) in zero (closed symbols) and an applied (open symbols) external magnetic field as a function of temperature. The solid curve shows the fitting of the data using Eq. (\ref{eq:Cs}) and the value of the superconducting gap can be estimated. Inset shows the plot of $C/T$ as a function of temperature near $T_c$. The data are linearly extrapolated around $T_c$ to determine the jump in the specific heat $\Delta C/T_c=(C_s-C_n)/T_c$. (c) The rescaled electronic specific heat in the superconducting state, plotted on the logarithmic scale, as a function of $T^{-1}$ for $x=0$ and $0.95$. The straight lines from $T_c/T=0.8^{-1}=1.25$ are guides to the eyes, with slopes $\alpha$ related to $\Delta$ via $\alpha=\Delta/k_BT_c$.}
\end{figure}

The specific heat jump at $T_c$ can be suppressed by magnetic field, thereby exposing the normal state. The low temperature specific heat in the normal state can be adequately described by the following model:
\begin{equation}
C_n(T)=\gamma T+\beta T^3 + \kappa \frac{e^{\Theta_E/T}}{(e^{\Theta_E/T}-1)^2}\left(\frac{\Theta_E}{T}\right)^2,
\label{eq:Cn}
\end{equation}
where the first term is the contribution from conduction electrons, the second and the third terms are due to phonons with dispersion relations $\omega\propto q$ (Debye) and $q$-independent $\omega$ (Einstein), respectively. $\Theta_E$ is the Einstein temperature \cite{Gopalbook}. In the inset to Fig. \ref{fig3}(a) where $C_n/T$ is plotted against $T^2$, the lowest temperature data follow a straight line and the value of $\gamma$ and $\beta$ could be extracted. The excess specific heat at higher temperatures is attributed to the Einstein term. Using this model, we successfully described the low temperature specific heat of all compositions studied (main panel of Fig. \ref{fig3}(a) and Fig. S6-S7 in Supplemental Material).
%%%%%%%%%%%%%%%%Figure 3

Next, we extract the parameters relevant to superconductivity. We take the midpoint of the specific heat anomalies at low temperatures as $T_c$. Following the scheme shown in the inset to Fig. \ref{fig3}(b), we estimated the normalized specific heat jump $\Delta C/\gamma T_c$ with the value of $\gamma$ obtained from the analysis of $C_n$ described above. In addition, we estimate the superconducting gap $\Delta$ by analysing the zero field heat capacity data well below $T_c$ using
\begin{equation}
C_s(T)=A\exp\left(-\frac{\Delta}{k_BT}\right)+\beta T^3 + \kappa \frac{e^{\Theta_E/T}}{(e^{\Theta_E/T}-1)^2}\left(\frac{\Theta_E}{T}\right)^2,
\label{eq:Cs}
\end{equation}
The phonon contributions from the the second and the third terms in $C_s$ are assumed to be identical to that in $C_n$. Hence, $A$ and $\Delta$ are the only adjustable parameters in the fitting. The usage of a single $s$-wave gap is justified based on previous studies, in which no evidence of low energy quasiparticle excitation was detected in the superconducting state, indicating a nodeless superconducting gap function \cite{Kase11,Hayamizu11,Wang2012,Zhou2012, Biswas14}. For $T\rightarrow T_c$, the BCS theory predicts that the gap closes in a fashion $\sim\sqrt{T_c-T}$ \cite{BCS1,BCS2}. Therefore, we fit the zero field heat capacity data below $0.8T_c$ with Eq. (\ref{eq:Cs}) for all seven samples (see Fig. \ref{fig3}(b) for $x=0.9$, and Fig. S2-S8 for others). The validity of our model is clearly demonstrated in Fig. \ref{fig3}(c) where the electronic part of the specific heat at zero field, $C_{es}$, is plotted on the logarithmic scale as a function of $T^{-1}$, with the solid lines drawn using the values of $\Delta$ extracted from the analysis of $C_s(T)$\cite{alphamodel}.

The results of these analyses are summarised in Fig. 4. In Fig. 4(a), we overlay $T^*$ and $T_c$ determined from the current heat capacity measurement on the universal phase diagram constructed earlier \cite{Goh14}. For $x>0.5$, the \Tstar\ anomaly becomes significantly smeared out in specific heat, although it is still visible in electrical resistivity. For all compositions studied here, $T_c$ values are in excellent agreement with previous reports \cite{Kase11, Goh14}. The phase diagram constructed clearly shows the suppression of \Tstar\ with increasing Ca content, giving rise to a structural QCP at $x=0.9$. A broad dome-like variation in $T_c$ is found to peak around the QCP.

We adopt the viewpoint that the second-order structural phase transition is accompanied by the softening of the relevant phonon mode frequency $\omega_{s,Q}$ on approaching \Tstar\ from $T>T^*$, and hence the branch of $\omega(q)$ on which this mode lies will be affected. In Landau's theory, $\omega_{s,Q}^2\propto(T-T^*)$ plays the role of the inverse generalised susceptibility \cite{Cowley1980, Dovebook}. Below \Tstar, the structure relaxes to a new stable configuration, and $\omega_{s,Q}$ will harden again. This temperature dependence of the soft phonon mode frequency across a structural transition has been most clearly observed in ferroelectric compounds, e.g. SrTiO$_3$ \cite{Cowley1969,Shirane1969,Fleury1968}. For the present system, $T^*=0$~K at the QCP. Consequently, at the QCP, $\omega_{s,Q}\rightarrow0$ at 0~K and we are left with an abundance of low-lying phonon modes at low temperatures.

First evidence of the existence of the soft mode comes from the electrical resistivity measurements of both \CaSrRhx\ and \CaSrIrx. At the structural QCP, namely in (Ca$_{0.9}$Sr$_{0.1}$)$_3$Rh$_4$Sn$_{13}$ at ambient pressure and Ca$_3$Ir$_4$Sn$_{13}$ under 18 kbar, a distinct linear-in-$T$ variation of the electrical resistivity was observed over a wide temperature range, which was attributed to a strong coupling to the soft phonon modes \cite{Klintberg2012, Goh14}.  Lattice dynamics calculations further revealed that the soft mode is located at M(0.5, 0.5, 0) \cite{Tompsett14, Goh14}, which is consistent with the structural refinement performed using data from x-ray diffraction studies \cite{Klintberg2012, Goh14}, and a recent inelastic scattering results in \CaIr \cite{Mazzone2015}. In \CaRh, which is beyond the structural QCP and hence does not undergo a structural transition, the soft mode frequency is $\omega_{s,Q}\sim0.35$~THz \cite{Goh14}. Since $\omega_{s,Q}$ is a local minimum at M, we can Taylor-expand around $\omega_{s,Q}$, and obtain $\omega\propto\sqrt{c^2q'^2+\omega_{s,Q}^2}$, where $q'$ is the wavevector measured from M. We emphasise that the lattice dynamics calculations were performed for 0~K. Thus, to induce a structural quantum phase transition from the quantum disorder side, one needs to reduce $\omega_{s,Q}$ to zero, resulting in a phonon dispersion relation $\omega\propto q'$. Hence, at the QCP, an additional, acoustic-like phonon branch emerges at M.
%%%%%%%%%%%%%%%%Figure 4
\begin{figure}[!t]\centering
       \resizebox{8.5cm}{!}{
              \includegraphics{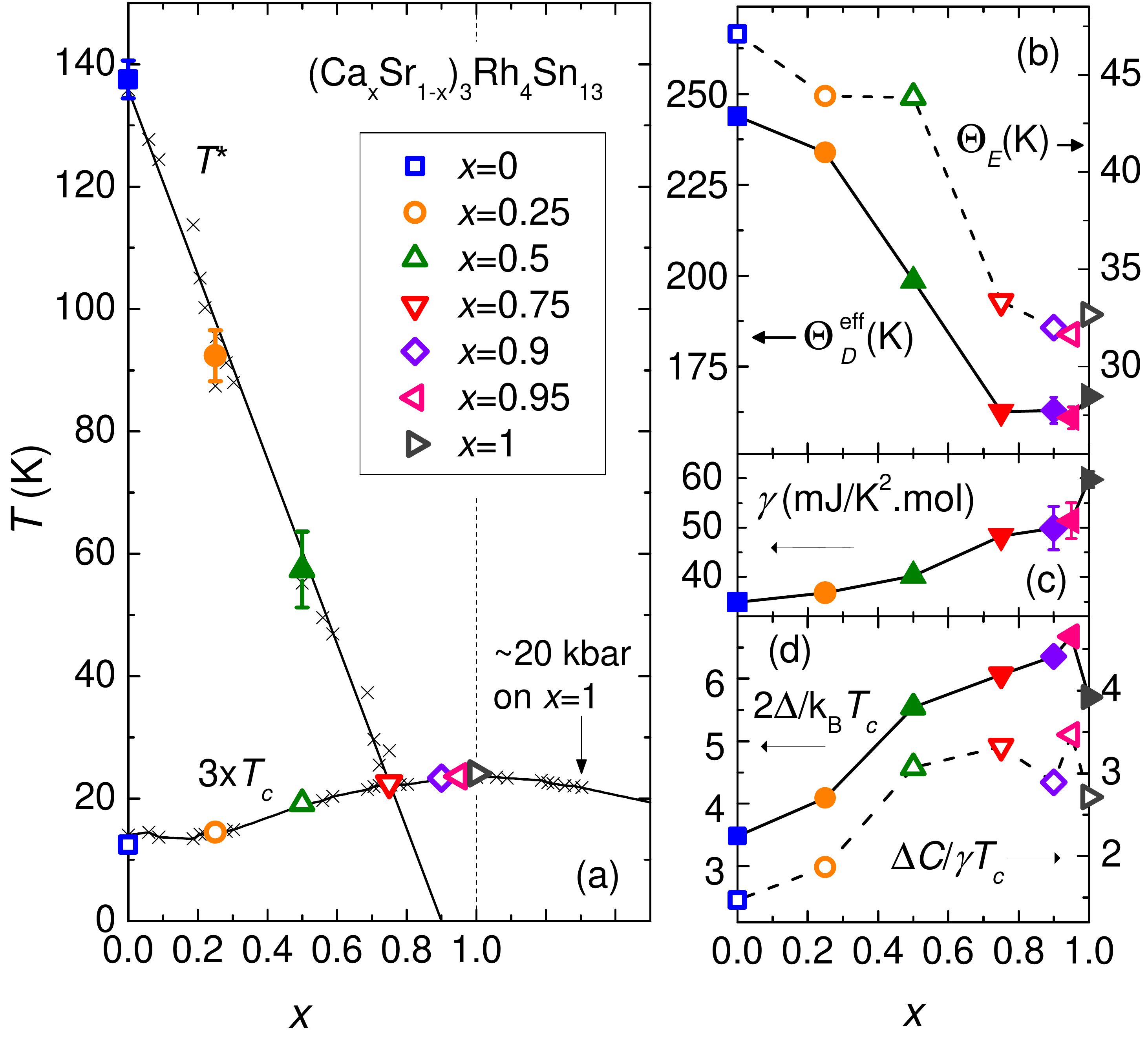}  }           				
              \caption{\label{figpd} (Color online) (a) The phase diagram of \CaSrRhx\ on the $T-x$ plane. Polygon symbols represent data extracted from heat capacity measurement. Cross symbols are data from electrical resistivity measurement in Ref. \cite{Goh14}. The region beyond $x=1$ is accessible by applying pressure. Right panels show the $x$ dependence of (b) the effective Debye temperature and the Einstein temperature, (c) $\gamma$, (d) $\Delta C/\gamma T_c$ and $2\Delta/k_B T_c$, respectively. The numerics of these parameters are tabulated in Table S1 of the Supplemental Material.}
\end{figure}
%%%%%%%%%%%%%%%%%%%%

We are now in the position to explain the $x$-dependence of the remaining quantities as shown in Fig. \ref{figpd}, and to put the results in the context of structural quantum criticality associated with this system. First of all, the contribution of the additional acoustic-like branch to the specific heat will be proportional to $T^3$, following Debye's treatment. Indeed, we found that $\beta$ (c.f. Eq. (1)) is significantly larger, or equivalently, the effective Debye temperature $\Theta_D^{\text{eff}}$ is considerably suppressed, at around $x=0.9$ \cite{effectivedebye}. Our analysis indicates a rapid decrease in $\Theta_D^{\text{eff}}$ as $x$ increases, with a broad minimum around $x=0.9$ where $\Theta_D^{\text{eff}}$ is $\sim 33\%$ lower than that at $x=0$. Concurrent with the strong variation of $\Theta_D^{\text{eff}}$, a qualitatively similar variation is observed in $\Theta_E$. This is consistent with our expectation that an abundance of low-energy phonon modes become readily populated as a result of the softening of $\omega_{s,Q}$.

%Hence, the evolution of $\Theta_D^{\text{eff}}$ and $\Theta_E$ are consistent with the existence of structural instability near $x=0.9$. On the contrary, $\gamma$ only exhibits a rather weak variation, with an average value of 39~mJ/K$^2$.mol and $\pm$10\% variation across the phase diagram.

From Fig. 4(d), we see that in addition to an enhancement in $T_c$ with increasing $x$, both $\Delta C/\gamma T_c$ and $2\Delta/k_BT_c$ are appreciably larger on the Ca-rich end of the phase diagram. Criticality in $\Delta C/\gamma T_c$ can be clearly observed near the structural QCP at $x=0.95$. Except for $x=0$, $\Delta C/\gamma T_c$ is noticeably larger than the BCS value of 1.43 \cite{BCS1, BCS2, Poolebook} for all samples studied, and reaches a value as high as $\sim$3.47 at $x=0.95$. Note that both $T_c$ and $\gamma$ increase with increasing $x$ (Figs. \ref{figpd}(a) and (c)), therefore the enhancement in $\Delta C/\gamma T_c$ is accompanied by a much larger $\Delta C$. Moreover, $2\Delta/k_BT_c$ also experiences a sizeable enhancement near the structural QCP, reaching a remarkably high value of $\sim$6.67 at $x=0.95$. In fact, theory of strong coupling superconductivity give $|\Delta C/\gamma T_c|_{max}\sim2.98$ and $|2\Delta/k_BT_c|_{max}\sim5.57$ \cite{Carbotte1990}, which are both smaller than our observed values near the structural QCP. Similar observation has recently been reported from $\mu$SR studies, which found a large $2\Delta/k_BT_c$ of $\sim8$ near the pressure-induced structural quantum critical point in the sister compound Ca$_3$Ir$_4$Sn$_{13}$ \cite{Biswas2015}.

Combining all parameters relevant to the normal and superconducting states from systematic studies across the phase diagram of \CaSrRhx, an intriguing picture emerges: at $x=0$, we have a superconductor close to the weak coupling limit, with both $2\Delta/k_BT_c$ and $\Delta C/\gamma T_c$  close to the expected ratios in the BCS theory \cite{BCS1, BCS2, Poolebook}. This is in stark contrast to the Ca-rich end, in the vicinity of the structural QCP, where these ratios are significantly enhanced. Thus, the phase diagram can be regarded as the outcome of fine tuning the coupling strength, which ultimately give rise to the intricate interplay between the phases observed.

In summary, we have analysed the heat capacity data of \CaSrRhx, in the normal and superconducting states, for seven $x$ values straddling across the structural QCP at $x=0.9$. The effective Debye temperature and the Einstein temperature are found to be significantly reduced near $x=0.9$, which we attributed to the softening of a phonon mode responsible for the structural transition.  Additionally, $T_c$ is found to peak near $x=0.9$, accompanied by a remarkable enhancement in $2\Delta/k_BT_c$ and $\Delta C/\gamma T_c$ far beyond the BCS values. \CaSrRhx\ thus serves as an unprecedented model system for studying the intricate interplay between structural instability and strong coupling superconductivity.

\begin{acknowledgments}{\bf Acknowledgement.} We thank Takasada Shibauchi, David Tompsett, Malte Grosche, Yuji Matsuda and Jeff Tallon for helpful discussion. We acknowledge funding support from the CUHK (Startup Grant, Direct Grant No. 4053071), UGC Hong Kong (ECS/24300214), Grants-in-Aid from MEXT (22350029 and 23550152), and Glasstone Bequest, Oxford.
\end{acknowledgments}

%\begin{eqnarray}
%	\xi_{ab}^2=\Phi_0/2\pi H_{c2}^c \\
%	\xi_{ab}\xi_c=\Phi_0/2\pi H_{c2}^{ab}
%\end{eqnarray}

%%%%%%%%%%%%%%%%%% BIBLIOGRAPHY USING BIBTEX%%%%%%%%%%%%%%%%%%%%
%\bibliographystyle{phjcp.bst}
%\bibliographystyle{h-physrev}
%\bibliography{AllRefs2.bib}

\begin{thebibliography}{34}
\bibitem{effectivedebye}
The notion of the `effective Debye temperature' stems from the fact that the contribution of the acoustic-like branch at M is taken into account in the conversion from $\beta$ to $\Theta_D^{\text{eff}}$, which is different from the original Debye picture where at most three acoustic branches are contributing.
\end{thebibliography}

%%%%%%%%%%%%%%%%%% END OF BIBLIOGRAPHY USING BIBTEX%%%%%%%%%%%%%%%%%%%%

\end{document}